\documentclass[aps,prl,twocolumn,showpacs,floatfix,nofootinbib]{revtex4}

\usepackage{psfrag,graphicx}
\usepackage{dcolumn}
\usepackage{amsmath,amssymb}
\usepackage{bm}

\newcommand{\be}{\begin{equation}}
\newcommand{\ee}{\end{equation}}
\newcommand{\bq}{\begin{eqnarray}}
\newcommand{\eq}{\end{eqnarray}}

\def\be{\begin{equation}}
\def\ee{\end{equation}}
\def\bea{\begin{eqnarray}}
\def\eea{\end{eqnarray}}

\def\1/2{\frac{1}{2}}

\begin{document}

\title{Quantum phases of electric dipole ensembles in atom chips}

\author{Jiannis K. Pachos}
\affiliation{Department of Applied Mathematics and Theoretical
Physics, University of Cambridge, Cambridge CB3 0WA, UK.}

\date{\today}

\begin{abstract}

We present how a phase factor is generated when an electric dipole
moves along a closed trajectory inside a magnetic field gradient. The
similarity of this situation with charged particles in a magnetic
field can be employed to simulate condensed matter models, such as the
quantum Hall effect and chiral spin Hamiltonians, with ultra cold
atoms integrated on atom chips. To illustrate this we consider a
triangular configuration of a two dimensional optical lattice, where
the chiral spin Hamiltonian $\vec{\sigma}_i \cdot \vec{\sigma}_j
\times \vec{\sigma}_k$ can be generated between any three neighbours
on a lattice yielding an experimentally implementable chiral ground
state.

\end{abstract}


\maketitle

The interaction of the electromagnetic field with charged particles
has unveiled a variety of geometrical and topological effects, such as
the Aharonov-Bohm effect \cite{Aharonov1}, and plays a predominant
role in the generation of collective phenomena, such as the quantum
Hall effect and fractional statistics \cite{Prange}. The Aharonov-Bohm
effect associates a quantum phase with the state of a charged
particle. This phase is proportional to the enclosed magnetic flux
when the particle undergoes a looping trajectory and it is obtained
even if the magnetic field is vanishing in the immediate
neighbourhood of the trajectory. This intriguing fact inspired a
series of extensions, e.g. the reciprocal or the dual of the original
Aharonov-Bohm effect \cite{Aharonov2,Wilkens,He,Spavieri}. Here, we
shall consider another possible generalisation that can be applied to
the neutral atom technology of Bose-Einstein condensates and provides a
unique laboratory for the simulation of condensed matter models,
e.g. chiral spin lattices or the quantum Hall effect, exhibiting
advanced controllability and long coherence times. Such
engineered systems can play a significant role in probing topological
quantum simulations and error-free quantum information processing
\cite{Kitaev}.

With advances in cold atom technology it has been possible to
simulate many solid state effects \cite{Kukl,Duan,Pachos} by employing
neutral atoms and techniques from quantum optics. Initially, it seemed
that the neutrality of the atoms significantly restricts the range of
phenomena this technology can simulate. Nevertheless, great effort
has been made to develop neutral atom techniques, generating suitable
phase factors by laser radiation to simulate the behaviour of charged
particles \cite{Jaksch,Jaksch1,Duan,Liu}. These include proposals for
the creation of the Hofstadter butterfly \cite{Jaksch1} and the
lattice implementation of the quantum Hall effect \cite{Sorensen}.
These realisations are limited in terms of their possible applications
and it still remains to develop a technique for the efficient
simulation of the effect of electromagnetic fields on charged spin
particles.

Here, an alternative approach is presented, which allows us to go
well beyond these applications. In the following we consider the electric
dipole moment of the atoms in the presence of an appropriate
electromagnetic field. The resulting effect is equivalent to a charged
particle moving in the presence of a magnetic field. This can simulate
both the continuous case of electrons confined in a two dimensional
plane, resulting in the quantum Hall effect \cite{Sen,Kalmeyer}, 
as well as the case of discrete lattice systems, which can result in the 
realisation of chiral states \cite{Wen}. The advantage of the 
present scheme is based 
on the simplicity of the required control procedure: utilising the 
atomic electric dipole as an additional degree of freedom provides 
adequate resources for probing even more complex structures, such as 
charged spin systems.

Let us first see the general setup for implementing interactions
between a particle and an external electromagnetic field. Consider
the case where a particle has a charge $e$, and an electric or a
magnetic moment, given by $\vec{d}_e$ or $\vec{d}_m$ respectively.
The minimal coupling of the particle with the electromagnetic
field is given by substituting its momentum for
\begin{equation}
\vec{p} \rightarrow \vec{p} +e \vec{A} +\vec{d}_m
\times \vec{E} +(\vec{d}_e \cdot
\vec{\nabla})\vec{A}, \label{mom}
\end{equation}
where $\vec{E}$ is the electric field and $\vec{A}$ is the magnetic
vector potential. The second term in (\ref{mom}) requires that the
particle is charged and gives rise to the well known Aharonov-Bohm
effect \cite{Aharonov1}. While its topological character with the
employment of a magnetic solenoid is of great conceptual value, and
has been verified experimentally \cite{Tonomura}, it has also
given rise to a variety of applications in the solid state arena
\cite{Yacoby}. The third term in (\ref{mom}) is the origin of the
Aharonov-Casher effect \cite{Aharonov2}, which is reciprocal to
the Aharonov-Bohm effect. It involves the circulation of a
magnetic dipole around a charged straight line and it has been
experimentally verified \cite{Cimino}. Recent experiments have
been performed that generalise the Aharonov-Casher effect thereby
partly overcoming original technological complications
\cite{Hinds}. The fourth term involves the coupling of the
electric dipole moment to a differential of the vector potential
and its consequences in cold atom technology will be the focus of
this Letter.

\begin{center}
\begin{figure}[!ht]
\resizebox{!}{2.3cm} {\includegraphics{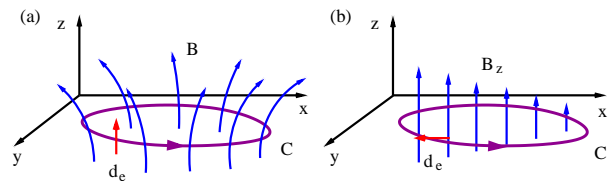} }
\caption{\label{zmag} The path circulation of the electric dipole
in the inhomogeneous magnetic field. Fig. (a) depicts magnetic
field lines sufficient to produce the appropriate non-vanishing
gradient of $B_z$ along the $z$ axis. Fig. (b) depicts only
$B_z$ and how it varies along the direction of the dipole,
$\vec{d}_e$.}
\end{figure}
\end{center}

In fact, electric dipoles can give rise to a variety of different
phenomena that generate quantum phases when they undergo a cyclic
trajectory $C$. Starting from (\ref{mom}) and by employing Stokes's
theorem the phase factor contribution to the final state of the
dipole is eventually given by
\begin{eqnarray}
\phi&=&
\int_S \big[\vec{\nabla} \times (\vec{B} \times \vec{d}_e) +
  \vec{\nabla} \times \vec{\nabla} (\vec{d}_e \cdot \vec{A}) \big]
\cdot d\vec{s},
\label{phase2}
\end{eqnarray}
where $S$ is a surface bordered by the cyclic path $C$ 
and $d\vec{s}$ is its elementary area. The
second term on the right hand side is the curl of the gradient of
$\vec{d}_e \cdot \vec{A}$ and can be taken to be zero, assuming
continuity. The case where $\vec{d}_e \cdot \vec{A}$ is a
multi-variable function, as might be produced when the dipole passes
through a magnetic sheet, results in topological effects that have
been studied in \cite{Spavieri}. Hence, from (\ref{phase2}), we
obtain
\begin{eqnarray}
\phi&=&
\int_S \big[ (\vec{d}_e \cdot \vec{\nabla}) \vec{B} - \vec{d}_e
  (\vec{\nabla} \cdot \vec{B}) \big] \cdot d\vec{s}.
\label{phase3}
\end{eqnarray}
The second term on the right hand side is zero as can be seen from the Maxwell
equations. Alternatively, a configuration with an infinite chain
of magnetic monopoles, that alter the usual Maxwell equations, can
lead to the generation of a topological phase dual to the
Aharonov-Casher one \cite{Wilkens,He}. For $\vec{\nabla} \cdot
\vec{B}=0$ we finally obtain
\begin{eqnarray}
\phi&=& \int_S (\vec{d}_e \cdot \vec{\nabla})\vec{B} \cdot
d\vec{s}. \label{phase4}
\end{eqnarray}
This equation can be viewed as the {\em differential Aharonov-Bohm
effect}. By inspection of relation (\ref{phase4}) we see that a
nontrivial phase can be produced if we generate an inhomogeneous
magnetic field in the neighbourhood of the dipole. In particular, a
non-zero gradient of the magnetic field component perpendicular to the
surface $S$, varying in the direction of the dipole, ensures a
non-zero phase factor. The dual effect with a magnetic dipole and a
gradient of an electric field can also give a similar phase factor
that originates from the third term in (\ref{mom}), while a homogeneous
electric field will just orient the electric dipole appropriately. As
a realisation of (\ref{phase4}), we can take $S$
to lie on the $x$-$y$ plane and $\vec{d}_e$ to be perpendicular to
the surface $S$. A non-zero phase, $\phi$, is produced if
there is a non-vanishing gradient of the magnetic field along the
$z$ direction (see Figure \ref{zmag}(a)). Alternatively, if
$\vec{d}_e$ is along the surface plane, then a non-zero phase is
produced if the $z$ component of the magnetic field has a
non-vanishing gradient along the direction of $\vec{d}_e$ as shown
in Fig. \ref{zmag}(b). Furthermore let us consider the case where 
the gradient of the magnetic field is spatially restricted, e.g. in a finite
interval along the $x$ direction. Then a trajectory of an electric
dipole outside this area, $S$, gives rise to a topological phase 
independent of the local details of the trajectory.
\vspace{-0.3cm}
\vspace{-0.7cm}
These configurations of magnetic field gradients have to satisfy
the Maxwell equations, $\vec{\nabla}\cdot \vec{B}=0$ and
$\vec{\nabla}\times \vec{B}=0$. Indeed, for $\partial B_z/
\partial x \neq 0$ one should allow for an additional
variation of the $x$ component of the magnetic field along the $z$
direction with equal gradient. It can be verified easily that this
satisfies Maxwell equations and provides the
nontrivial phase factor given in (\ref{phase4}). The same holds in the
presence of a uniform electric field used to orient the
electric dipole appropriately.

Let us now consider the implementation of the phase factor (\ref{phase4})
with neutral atoms confined on a two dimensional plane. The electric
dipole moment of an atom, which can be induced with the application of an
external homogeneous electric field, is of the order $|\vec{d}_e|
\approx n^2 e a_0$, where $n$ is the principal quantum number of the
electron orbit, $e$ is its charge and $a_0$ is the Bohr radius. It is
even plausible to consider the employment of Rydberg atoms for which
$n$ is quite large \cite{Jaksch2}. By applying an external macroscopic
magnetic field we achieve gradients of the order
$\partial B_z / \partial x \approx 10^2$-$10^3$ Gauss/m. Much
larger gradients of the magnetic field are feasible with atom
chip technology due to the miniaturisation of their structure. For
example, gradients of the order of $10^8$ Gauss/m have already
been reported at Caltech for the manipulation of ultra cold 
neutral atoms \cite{Lev}. This technology should prove very fruitful
for the future of Bose-Einstein experiments where whole atomic
ensembles can be mounted on atom chips \cite{Dekker,Folman,Horak}. 
With these at hand we can
calculate that for atomic number $n=1$ and for the circulation of
an area of the order of the optical wavelength ($\sim 1\mu$m) squared
one can achieve phases of the order of $\phi \approx 8 > 2\pi$.
Hence, by varying the gradient of the magnetic field or even the
orientation of the electric dipole via an additional electric
field it is possible to obtain an arbitrary value for the phase
factor $e^{i\phi}$. This provides a great degree of controllability
when simulating charged particles in the presence of magnetic
fields of arbitrarily large intensity. Next we shall see
applications of this property.

An important tool in the manipulation of atomic ensembles is the
employment of optical lattices. They can generate one, two or three
dimensional structures of potential minima for the atoms. In
particular, a one dimensional optical lattice is sufficient to confine
the atomic ensemble in parallel planes above an atom chip, providing
two dimensional confinement, a necessary condition for the
implementation of the proposed phase factor (\ref{phase4}) with atomic
ensembles. For a sufficiently high flux of the effective magnetic
field through an area $S$, given by $e^* B^* S / h$, and low
two-dimensional densities of atoms, $n_{2D}$, one can obtain
\cite{Cooper}, for example, the fractional filling factor $\nu =
n_{2D} / (e^*B^*/h)=1/2$. This corresponds to the fractional quantum
Hall effect \cite{Wen,Kalmeyer} which can be described by the $m=2$
Laughlin wavefunction \cite{Laughlin}.

Moreover, additional optical lattices can be applied to create a
regular structure on this plane in order to simulate two
dimensional topological spin effects. This is achieved by
generating complex tunneling interactions along the planar lattice
sites. In particular, consider two species of atoms, namely $\sigma =
\{\uparrow,\downarrow\}$, that are superimposed with particular
configurations of optical lattices. The evolution of the system, for
atoms restricted in the lowest Bloch-band, is described by the
Bose-Hubbard Hamiltonian that is comprised of tunneling transitions of
atoms between neighbouring sites of the lattice, $V=-\sum_{i\sigma}
(J^\sigma_i a_{i\sigma}^\dagger a_{i+1 \sigma} +\text{H.c.})$, and
collisional interactions between atoms in the same site,
$H^{(0)}=\frac{1}{2} \sum _{i \sigma \sigma'} U_{\sigma \sigma'}
a^{\dagger}_{i\sigma}a^{\dagger}_{i\sigma'}a_{i\sigma'}a_{i\sigma}$
with couplings, $U_{\uparrow \uparrow}$, $U_{\downarrow
\downarrow}$ and $U_{\uparrow \downarrow}$. By arranging for the
tunneling couplings to be small with respect to the collisional ones,
the system can be brought into the Mott insulator phase with only one
atom per lattice site \cite{Raithel}. Hence, one can assume
that the space of states of the system is spanned by the basis
states $|\uparrow\rangle_i \equiv |n^\uparrow_i=1,
n^\downarrow_i=0\rangle$ and $|\downarrow\rangle_i \equiv
|n^\uparrow_i=0, n^\downarrow_i=1 \rangle$, where $n^\sigma_i$ is
the number of atoms of species $\sigma $ in site $i$.

In the rotated frame with respect to $H^{(0)}$ one can expand the
total Hamiltonian in terms of the small tunneling interactions,
obtaining an effective Hamiltonian that describes the tunneling of
atoms without populating energetically unfavourable states that have
two or more atoms per site. Subsequently, this Hamiltonian can be
expressed in terms of the Pauli operators. Such expansion has provided
a variety of two and three spin Hamiltonians
\cite{Kukl,Duan,Pachos}. By the additional employment of a magnetic
field gradient and by considering the electric dipole of the atoms, it
is possible to generate complex tunneling couplings of the form
$J=e^{i\phi}|J|$ with
\begin{equation}
\phi=\int_{\vec{x}_i}^{\vec{x}_{i+1}}(\vec{d}_e \cdot
\vec{\nabla})\vec{A} \cdot d \vec{x}. \nonumber
\end{equation}
Here $\vec{x}_i$ and $\vec{x}_{i+1}$ denote the positions of the
lattice sites connected by the tunneling coupling $J$. This is
equivalent to the tunneling of a charged particle (e.g. electron)
along a lattice in the presence of a magnetic field.

In order to isolate the new effects generated by the consideration
of complex tunneling couplings, we restrict ourselves to purely
imaginary ones, i.e. $J_i^{\sigma}= \pm i |J_i^{\sigma}|$. We
also focus initially on the case where the optical lattices
generate a two dimensional structure of equilateral triangles. Such a
non-bipartite structure is
necessary in order to manifest the breaking of the symmetry under
time reversal, $T$, in our model, eventually producing an effective
Hamiltonian that is not invariant under complex conjugation of the
tunneling couplings. Moreover, as the second order perturbation
theory is manifestly $T$ symmetric, we need to consider the third
order.
In that case the effective Hamiltonian becomes
\begin{equation}
\begin{split}
H_{\text{eff}}&=\sum_{\langle i,j \rangle} \Big[ \tau^{(1)}\sigma^z_i\sigma^z_j
+ \tau^{(2)}(\sigma^x_i\sigma^x_j+ \sigma^y_i\sigma^y_j)+ 
\\& \tau^{(3)}(\sigma^x_i\sigma^y_j- \sigma^y_i\sigma^x_j)\Big]+
\tau^{(4)}\sum_{\langle i,j,k \rangle}\vec{\sigma}_i\cdot \vec{\sigma}_j
\times \vec{\sigma}_k,
\end{split}
\label{complexboson}
\end{equation}
with $\vec{\sigma}=(\sigma^x,\sigma^y,\sigma^z)$ and $\langle
...\rangle$ denoting nearest neighbour sites. The couplings
appearing in (\ref{complexboson}) are given by
\begin{equation}
\begin{split}
&\tau^{(1)} = {J^\uparrow}^2( {1\over U_{\uparrow\uparrow}} -{ 1 \over 2
  U_{\uparrow \downarrow}})+ (\uparrow \leftrightarrow
  \downarrow),\,\,\,
\tau^{(2)}={J^\uparrow J^\downarrow \over
  U_{\uparrow\downarrow}},\\
&\tau^{(3)}= i{{J^\uparrow}^2 J^\downarrow \over
  U_{\uparrow\uparrow}}\Big( {1 \over 2 U_{\uparrow\uparrow}} + {1
  \over U_{\uparrow\downarrow}} \Big) + (\uparrow \leftrightarrow
  \downarrow),\,\,\,
\\&
\tau^{(4)}= i{{J^\uparrow}^2 J^\downarrow \over
  U_{\uparrow\uparrow}}\Big( {1 \over 2 U_{\uparrow\uparrow}} + {1
  \over U_{\uparrow\downarrow}} \Big) - (\uparrow \leftrightarrow
  \downarrow). \nonumber
\end{split}
\end{equation}
Extraneous Zeeman terms in the $z$ direction can be dealt with by Raman
transitions that effectively create a compensating field of the form
$B_z= 2({{J^\uparrow}^2 / U_{\uparrow\uparrow}} -{{J^\downarrow}^2
  / U_{\downarrow\downarrow}})$. The coupling constants, $\tau^{(i)}$, 
can take various values. For example, one can choose
$U_{\uparrow \downarrow} \rightarrow \infty$
(e.g. by employing fermionic atoms) so that $\tau^{(2)}$ vanishes. 
For certain values of the collisional couplings one can also vary
the tunneling couplings such that $\tau^{(1)}=0$. Moreover, by 
appropriately tuning the phase
of the tunneling couplings one can choose either
$\tau^{(3)}$ or $\tau^{(4)}$ to vanish.

Remarkably, with this physical proposal, a chiral three-spin
interaction appears in (\ref{complexboson}), which can be isolated,
especially from the Zeeman terms that are predominant in equivalent
solid state systems. This interaction term is also known in
the literature as the {\em chirality operator} \cite{Wen}. It breaks
the time reversal symmetry of the system as a consequence of the
externally applied field. A chiral spin state is then a state for
which the expectation value of the chirality operator has a nonzero
value independent of the position of the plaquette $\langle i,j,k
\rangle$.
\begin{center}
\begin{figure}[!ht]
\resizebox{!}{7.5cm} {\includegraphics{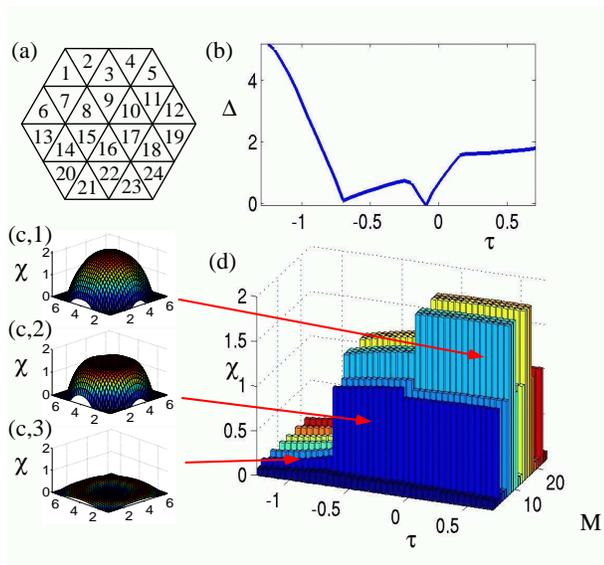} }
\caption{\label{xxyyboundaries} (a) The hexagonal structure with 19 spins
  on the vertices and the 24 plaquettes. (b) The energy gap, $\Delta$, from the ground
  state to the excited one as a function of the
  coupling $\tau$. (c) The chirality, $\chi$, as a function of the
  plaquette positioned on the plane for three different couplings
  $\tau$ corresponding to (d), where the chirality, $\chi$, of each
  plaquette is given as a function of the coupling $\tau$ and the
  particular plaquette $M$.}
\end{figure}
\end{center}
\vspace{-0.5cm}

As a particular example, for studying the behaviour of the three spin
interaction term, we take a hexagonal configuration of 19 spin (see
Fig. \ref{xxyyboundaries}(a)). On this lattice we simulate Hamiltonian
(\ref{complexboson}) for $\tau^{(1)}=0$, $\tau^{(2)}=\tau\cdot \tau^{(4)}$ 
($\tau^{(4)}>0$) and $\tau^{(3)}=0$ with the additional
condition that the spins on the boundary experience
a strong magnetic field oriented in the z-direction. A numerical
simulation has been performed to obtain the energy gap, $\Delta$,
above the ground state (see Fig. \ref{xxyyboundaries}(b)) as well 
as the chirality of the ground state,
$\chi \equiv \langle \vec{\sigma}_i\cdot \vec{\sigma}_j \times
\vec{\sigma}_k  \rangle$, for each triangular plaquette of
neighbouring sites $i$, $j$ and $k$ (see Fig. \ref{xxyyboundaries}(c) 
and (d)). Indeed, Fig. \ref{xxyyboundaries}(b) shows that
there is criticality behaviour for $\tau_{c_1} \approx -0.7$ and \
$\tau_{c_2} \approx -0.1$, where the energy gap becomes zero. In Fig.
\ref{xxyyboundaries}(d) the chirality for different values of $\tau$
and for different plaquettes on the plane numbered by $M$ is
displayed where a reconstruction of the chirality on the plane is
given in \ref{xxyyboundaries}(c,1), (c,2) and (c,3) for the three
distinctive areas.

For $\tau<\tau_{c_1}$ the exchange interaction, $H_{\text{e}}=
\sum_{\langle ij \rangle}(\sigma^x_i\sigma^x_j+
\sigma^y_i\sigma^y_j)$, dominates the chiral interaction,
$H_{\text{c}}=\sum_{\langle ijk \rangle} \vec{\sigma}_i\cdot
\vec{\sigma}_j \times \vec{\sigma}_k$, forcing the spins to be
aligned, pointing upwards in agreement to the boundary conditions. In
this case $\chi$ is almost zero (see Fig.
\ref{xxyyboundaries}(c,3)). This holds also when we consider the Ising
or a Zeeman interaction dominating $H_{\text{c}}$. For
$\tau_{c_1}<\tau$, the interaction $H_{\text{c}}$ dominates giving a
non-zero value to $\chi$. At $\tau_{c_1}$ and $\tau_{c_2}$ the chiral
order parameter jumps indicating a quantum phase transition towards a
chiral phase of the spin system. In particular, the plaquettes that
have two sites on the boundary have small $\chi$ as both of the
boundary spins tend to have parallel orientation. There is a small
jump in $\chi$ around $\tau_{c_2}$ as $H_{\text{e}}$ induces a
transition between energetically favourable states from ferromagnetic
($\tau<0$) to antiferromagnetic ones ($\tau>0$), for which case
$\chi$ increases due to frustration. As can be easily calculated for
a single triangle, $H_{\text{e}}$ has doubly degenerate ground
states. This degeneracy is actually lifted by the chiral interaction
giving a non-zero $\chi$ for $\tau>0$. Indeed, one finds that on a
triangle the common ground state of the two interactions is $|\Psi
\rangle = {1 \over \sqrt{3}} \big(|\uparrow \uparrow \downarrow
\rangle + \omega |\uparrow \downarrow \uparrow \rangle + \omega^2
|\downarrow \uparrow \uparrow \rangle \big)$ which has non-vanishing
chirality. This fact is in agreement with the persistence of chirality
for large $\tau$ as shown in Fig. \ref{xxyyboundaries}(d), which is
the regime of the adopted perturbation theory. This experimentally
feasible domain of couplings exhibits as a ground state a chiral spin
state that may be possible to detect with the state of the art
technology.

This work was supported by the Royal Society.

\end{document}